\documentstyle[iopfts]{ioplppt}

\begin{document}

\title{A new measure of nonclassical distance}[A new measure 
of nonclassical distance] 

\author{Horia Scutaru}
\address{Department of Theoretical Physics, Institute of
Atomic Physics, PO BOX MG-6,
R-76900 Bucharest-Magurele, Romania\\
E-mail: scutaru@theor1.theory.nipne.ro}


\begin{abstract}
In the present paper we shall propose a new measure of the
nonclassical distance \cite{hill1}.
The proposed modification is based on the following
considerations. If $\rho_{1}$ and $\rho_{2}$ are density
operators, and $F(\rho_{1},\rho_{2})$ is the corresponding
fidelity, then from the  
inequalities \cite{fuchs}
\begin{eqnarray}
&&
\nonumber
2(1-\sqrt{F(\rho_{1},\rho_{2})}) \leq ||\rho_{1}-\rho_{2}||_{1}
\leq 2[1-(F(\rho_{1},\rho_{2}))]^{{1 \over 2}}
\end{eqnarray}
it is evident that the quantity 
\begin{eqnarray}
&&
\nonumber
\phi(\rho)= \sup_{\rho_{cl}}F(\rho_{cl},\rho)
\end{eqnarray}
can be used in the same extent as a measure of the
distance of the state $\rho$ to the set of classical
states $\rho_{cl}$ as the Hillery measure 
$\delta(\rho)=\sup_{\rho_{cl}}||\rho_{1}-\rho_{2}||_{1}$ \cite{hill1}.
$\phi(\rho_{cl})=1$ for any classical state and $\phi(\Gamma(\rho))
\geq \phi(\rho)$ if the map $\Gamma$ is the Gaussian noise map
\cite{hall,mussli}.
\end {abstract}

\maketitle

\section{Introduction}
In \cite{hill1}, Hillery gave a definition of 
the nonclassical distance
of radiation in terms of the trace norm as
\begin{equation}
\delta(\rho)= \inf_{\rho_{cl}}||\rho - \rho_{cl}||,
\end{equation}
where $\rho$ is the density matrix of the nonclassical radiation
and $\rho_{cl}$ is that of an arbitrary classical field, while
$||A||_{1}$ is the trace norm of the operator $A$.
The Hillery noncassical distance has the following properties:
\begin{enumerate}
\item $\delta(V(u)\rho V(u)^{\dagger})=\delta(\rho)$;
\item $ 0 \leq \delta(\rho) \leq 2$;
\item $\delta(\theta \rho_{1}+ (1-\theta)\rho_{2}) \leq 
\theta\delta(\rho_{1}) + (1-\theta) \delta(\rho_{2})$.
\end{enumerate} 
where the unitary operators $V(u)$ are the well known Weyl operators
giving a projective unitary representation of the vector group
${\bf R}^{2n}$ (see the next section).
This measure seems to be quite universal. Unfortunately, it is not
easy to use this measure in actual calculations. In practice, one
can only provide the upper bound and lower bound of the measure.

It is possible to define the distance between two quantum states
described by density operators in many ways \cite{wun,knorlo,monge,dod}. 
If the distance is small
these two density operators can be considered very similar to
each other. 
On the other hand a large distance means very different
density operators.

Another, more physical point of wiew is that presented in \cite{fuchs}. 
According to this point of view "the only physical
means available with which to distinguish two quantum states is that
specified by the general notion of quantum mechanical measurement". 
Instead of a metrical point of view a statistical point of view
is taken into account. A measurement being necessarily indeterministic 
and statistic the more physical measures of distance between two quantum
states are those which are based on the statistical-hypothesis
testing procedures. 

In the present paper we shall propose a new measure of the
nonclassical distance \cite{hill1}.
The proposed modification is based on the following
considerations. If $\rho_{1}$ and $\rho_{2}$ are density
operators, and $F(\rho_{1},\rho_{2})$ is the corresponding
fidelity, then from the  
inequalities \cite{fuchs}
\begin{equation}
2(1-\sqrt{F(\rho_{1},\rho_{2})}) \leq ||\rho_{1}-\rho_{2}||_{1}
\leq 2[1-(F(\rho_{1},\rho_{2}))]^{{1 \over 2}}
\end{equation}
it is evident that the quantity 
\begin{equation}
\phi(\rho)= \sup_{\rho_{cl}}F(\rho_{cl},\rho)
\end{equation}
can be used in the same extent as a measure of the
distance of the state $\rho$ to the set of classical
states $\rho_{cl}$ as Hillery's measure 
$\delta(\rho)$.

Let $\rho_1$ and $\rho_2$ be two density operators which describe 
two mixed states. The transition probability $P(\rho_1,\rho_2)$ 
has to satisfy the following natural axioms:

\begin{enumerate}
\item $P(\rho_1,\rho_2) \leq 1$ and $P(\rho_1,\rho_2) = 1$ 
if and only if $\rho_1 = \rho_2$;

\item  $P(\rho_1,\rho_2) = P(\rho_2,\rho_1)$;

\item  If $\rho_1$ is a pure state, $\rho_1 = |\psi_1><\psi_1|$ then 
\newline $P(\rho_1, \rho_2) = <\psi_1|\rho_2|\psi_1>$;

\item  $P(\rho_1, \rho_2)$ is invariant under unitary transformations on the 
state space;

\item  $P\left(\rho_1|_{\cal A},
 \rho_2|_{\cal A}
\right) \geq P(\rho_1, \rho_2)$ for any complete subalgebra 
of observables ${\cal A}$;

\item $P(\rho_1\otimes\sigma_1, \rho_2\otimes\sigma_2 ) = 
P(\rho_1, \rho_2)P(\sigma_1, \sigma_2)$.

\item $P( \mu_{1} \rho_{1} + \mu_{2} \rho_{2}, \sigma ) \geq
\mu_{1} P( \rho_{1} , \sigma )+ \mu_{2} P( \rho_{2} , \sigma )$
when $0 \leq \mu_{1},\mu_{2} \leq 1$, $\mu_{1} + \mu_{2} =1$.  
\end{enumerate}

Uhlmann's transition probability for mixed states \cite{bur,uhl,jos,bart}
\begin{equation}
P(\rho_1, \rho_2) = 
\left[Tr\left(\sqrt{\rho_1}\rho_2\sqrt{\rho_1}\right)^{1/2}\right]^2\label{2}
\end{equation}
satisfies properties 1--7. The fidelity is defined by 
$F(\rho_1, \rho_2) = P(\rho_1, \rho_2)$. A detailed analysis for the 
structure of the transition probability was hampered by the factors 
containing square roots.
Due to technical difficulties in the computation of fidelities, 
few concrete examples of analytic calculations are known. 
The first results in an infinite-dimensional Hilbert 
space were recently obtained by
Twamley \cite{tw} for the fidelity of two thermal squeezed states and by 
Paraoanu and Scutaru \cite{pa} for the case of two displaced thermal states. 
In \cite{scut1} Scutaru has developed another calculation method which allowed 
getting the result for the case of two displaced thermal squeezed states
in a coordinate-independent form.
A general formula
for the fidelity of any two mixed Gaussian states (i.e. multimode
displaced thermal states 
\cite{feacol,cha,lo,hol1,oz,scut2}, 
from which the previous results
can be obtained as particular cases, has been obtained recently 
\cite{parscut}. 

Another modification is based on the following
considerations \cite{hol2}. If $\rho_{1}$ and $\rho_{2}$ are density
operators, then it is easy to establish \cite{hol2} the 
inequalities
\begin{equation}
2(1-Tr\sqrt{\rho_{1}}\sqrt{\rho_{2}}) \leq ||\rho_{1}-\rho_{2}||_{1}
\leq 2[1-(Tr\sqrt{\rho_{1}}\sqrt{\rho_{2}})^2]^{{1 \over 2}}
\end{equation}
from which it is evident that the quantity 
\begin{equation}
\chi(\rho)= \sup_{\rho_{cl}}Tr\sqrt{\rho_{cl}}\sqrt{\rho}
\end{equation}
can be used in the same extent as a measure of the
distance of the state $\rho$ to the set of classical
states $\rho_{cl}$ as the Hillery measure $\delta(\rho)$.
In the same time it is more easy to calculate this cantity
than Hillery's measure. When $\rho_{1}$ and $\rho_{2}$
are the density operators of multimode displaced squeezed
thermal states then the quantity $Tr\sqrt{\rho_{1}}\sqrt{\rho_{2}}$
ca be computed in an explicit way \cite{hol2}.

In the following we shall take as a fundamental test for a
good nonclassical distance the fact that it must increase under the action of
a Gaussian (or thermal) noise 
\cite{hall,mussli,glau,lac,vour,vouwer,fea,lou,vour1,lee,buz,haro,pmtm,dod1}. 
A drawback of the measure of the nonclassical distance which
is based on the Holevo overlap is given by the fact
that in this case the condition is not fullfiled.

\section{Multimode thermal squeezed states}

Let $(E, \sigma)$ be a phase space i.e. a vector space with
a symplectic structure $\sigma$. Then the commutation relations  
on $(E, \sigma)$ acting in a Hilbert space ${\cal H}$ are defined 
by a continuous family of unitary operators $\{ V(u), u \in E \}$
on ${\cal H}$ which satisfy the Weyl relations \cite{hol1,scut2}:
\begin{equation}
V(u)V(v)=\exp{{i \over 2} \sigma(u,v)}~V(u+v).
\end{equation}
Hence the family 
$\{V(tu), - \infty < t < \infty \}$ for a fixed
$u \in E$
is a group of unitary operators.

Then by the Stone theorem 
\begin {equation}
V(u) = \exp{i R(u)},
\end{equation}
where $R(u)$ is a selfadjoint operator.
From the Weyl relations we have
\begin{eqnarray}
&&
\nonumber
\exp{itR(u)}\exp{isR(v)}=\\
&&
\nonumber
\exp{its\sigma(u,v)}\exp{isR(v)}\exp{itR(u)}.\\
&&
\nonumber
\end{eqnarray}
By differentiation and taking $t=s=0$ one obtains
\begin{equation}
[R(u),R(v)]= -i\sigma(u,v)I.
\end{equation}
The operators $\{R(u), u \in E\}$ are called
canonical observables.
\newline The phase space $E$ is of even real dimension $2n$ 
and there exist in $E$
symplectic bases of vectors $\{e_j, f_j\}_{j=1,...,n}$, i.e.
reference systems such that $\sigma(e_j,e_k) = \sigma(f_j,f_k)=0$
and $\sigma(e_j,f_k) = - \sigma(f_k,e_j) =
\delta_{jk}$, $j,k=1,...,n$.
The coordinates $(\xi^j,\eta^j)$ of a vector $u\in E$ in a
symplectic basis $( u= \sum_{j=1}^n(\xi^je_{j}+\eta^jf_{j}) )$ are
called symplectic coordinates. The well known coordinate and momentum
operators are defined by $Q_{k}=R(f_{k})$ and $P_{k}=R(e_{k})$ for
$k=1,2,...,n$. Then the canonical observables $R(u)$ are linear
combinations of the above defined coordinate and momentum operators:
$R(u)= \sum_{j=1}^n(\xi^jP_{j}+\eta^jQ_{j}) )$.
 
There is a one-to-one
correspondence between the symplectic bases and the linear
operators $J$ on $E$ defined by $Je_k=-f_k $ and $Jf_k=e_k$,
$k=1,...,n$. The essential properties of these operators are:
$\sigma(Ju,u)\geq 0$, $ \sigma(Ju,v)+\sigma(u,Jv)=0$ ($u,v \in E$ and
$J^2=-I$, $I$ denotes the identity operator on $E$).Such operators
are called complex structures.
In the following
we shall use the matricial notations with $u\in E$ as column vectors.
Then $\sigma(u,v)=u^TJv$ and the scalar product is given by 
$\sigma(Ju,v)=
u^Tv, u,v\in E$. A linear operator $S$ on $E$ is called a
symplectic operator if $S^TJS=J$. When $S$ is a symplectic operator
then $S^T$ and $S^{-1}$ are also symplectic operators. The
group of all symplectic operators $Sp(E,\sigma)$ is called the
symplectic group of $(E,\sigma)$. The Lie algebra of $Sp(E,\sigma)$
is denoted by $sp(E,\sigma)$ and its elements are operators $R$
on $E$ with the property: $(JR)^T=JR$. Hence an operator $R$ on $E$
belongs to $Sp(E,\sigma) \cap sp(E,\sigma)$ iff $R^2=-I$.
If $J$ and $K$ are two complex
structures, there exists a symplectic transformation $S$
such that $J = S^{-1}KS$. 
For any symplectic operator $S$ we can define a new system
of Weyl operators $\{V(Su); u \in E\}$. Then from a well known
result on the unicity of the the systems of Weyl operator up to
a unitary equivalence it follows that there exists a unitary
operator $U(S)$ on ${\cal H}$ such that $V(Su)=U(S)^{\dag}V(u)U(S)$.

For any nuclear operator $O$ on ${\cal H}$ one defines the
characteristic function

\begin{equation}
CF_{u}(O)= Tr O V(u), ~~~~~u \in E.
\end{equation}
We give the properties of the characteristic function
which are important in the following \cite{hol1}:

\begin{enumerate}
\item $CF_{0}(O)=TrO$;
\item $CF_{u}\left[V(v)^{\dag}OV(v)\right]=
CF_{u}\left[O\exp{i\sigma(v,u)}\right]$;
\item $CF_{u}(O_{1}O_{2})=
\newline {1 \over (2 \pi)^{n}} \int CF_{v}(O_{1})
CF_{u-v}(O_{2})\exp{{i \over 2} \sigma(v,u)} dv$;
\item $CF_{Su}(O)=CF_{u}(U(S)OU(S)^{\dag})$.
\end{enumerate}

The multimode thermal squeezed states are defined
by the density operators $\rho$ whose characteristic functions
are Gaussians \cite{scut1,hol1,scut2}
\begin{equation}
CF_{u}(\rho)= \exp\left\{-{1 \over 4}u^{T}Au\right\}.
\end{equation}
where $A$ is a $2n \times 2n$ positive definite matrix,
called correlation matrix. From the last property of the
characterisic function, enumerated above, it follows that:
\begin{equation}
A_{U(S) \rho U(S)^{\dag}}=S^TA_{\rho}S
\end{equation}

Because the correlation matrix A is positive definite it
follows \cite{scut2,fol} 
that there exists $S \in Sp(E,\sigma)$ such that
\begin{equation} 
A = S^{T}{\cal D}S 
\end{equation}
where
${\cal D}=
\left(\matrix{D&0\cr 0&D\cr}\right)$
and $D \geq I$ is a diagonal $n\times n$ matrix.
   The most general real symplectic transformation $S\in Sp(E,\sigma)$
has \cite{scut2,bal}
the following structure:
\begin{equation}
S = O{\cal M}O^{'} 
\end{equation}
where
\begin{equation}
{\cal M}=\left(\matrix{M&0\cr 0&M^{-1}\cr}\right)
\end{equation}
and $O$, $O^{'}$ are symplectic and orthogonal $(O^{T}O=I)$
operators, and where $M$ is a diagonal $n\times n$ matrix.
As a consequence the most general form of a correlation matrix $A$ 
is given by:  
\begin{eqnarray}
A = O^{'T}{\cal M}O^{T}{\cal D} O{\cal M}O^{'} 
\end{eqnarray}
Various particular kinds of such matrices are obtained taking $O$,
$O^{'}$, ${\cal D}$ or ${\cal M}$ to be equal or proportional to 
the corresponding 
identity operator.
A pure squeezed state is obtained when ${\cal D}=I$. If this condition
is not satisfied, the state is a mixed state called thermal squeezed
state \cite{eza}. When ${\cal M}=I$ there is no squeezing and the 
correspondig states
are pure coherent states or thermal coherent states. All these states
have correlations between the different modes produced by the
orthogonal symplectic operators $O$ and $O^{'}$.

From the property 3 of the characteristic function
we have for two density operators $\rho_{1}$ an
$\rho_{2}$
\begin{eqnarray}
&&
\nonumber
CF_{u}(\rho_{1}\rho_{2})=
\left[det\left({A_{1}+A_{2} \over 2}\right)^{-{1 \over 2}}\right]\\
&&
\nonumber
\exp{\left\{-{1 \over 4}u^{T}\left[A_{2}-
(A_{2}-iJ)(A_{1}+A_{2})^{-1}(A_{2}+iJ)\right]u\right\}}.
\end{eqnarray}
When $\rho_{1}=\rho_{2}$ we have
\begin{equation}
CF_{u}(\rho^2)= (detA)^{-{1 \over 2}}\exp
\left\{-{1 \over 4}u^T\left({A-JA^{-1}J \over 2}\right)u\right\}.
\end{equation}
A state $\rho$ is pure iff $\rho^2=\rho$.
Then from the equality 
$CF_{u}(\rho^2)=CF_{u}(\rho)$ 
it follows that a Gaussian
state is pure iff
\begin{equation}
A=-JA^{-1}J,
\end{equation}
i.e. a Gaussian state is pure iff $JA \in Sp(E,\sigma)$.
Analogously, for a mixed state $\rho^2 < \rho$.
Then $CF_{u}(\rho^2) < CF_{u}(\rho)$ and as 
a consequence ${A-JA^{-1}J \over 2} > A$. 
Hence for any Gaussian state the correlation
matrix $A$ must satisfy the folloving restriction \cite{scut2}
\begin{equation}
A \leq -JA^{-1}J.
\end{equation}

\section{The classical Gaussian states}

A multimode squeezed thermal state is a classic\-al state
when it has a $P$-representa\-tion.
The $P$-distribution on the phase space which describes
such a state is the symplectic Fourier transform of the
normal ordered characteristic function 
\begin{equation}
CF^{N}_{u}(\rho) = \exp\{-{1 \over 4}u^T(A-I)u\}.
\end{equation}
The necessary and sufficient condition for the
existence of the symplectic Fourier transform
is the positive definiteness of the matrix $A-I$.
Then one has
\begin{eqnarray}
&&
\nonumber
P(v)=\\
&&
\nonumber
\pi^{-n}(\sqrt{det(A-I)})^{-1} \exp\{{1 \over 4}v^TJ(A-I)^{-1}Jv\}.\\
\end{eqnarray}

\section{The Gaussian noise}

One form of noise which has been extensively studied is the
thermal noise. The admixture of the thermal noise is described
by the semigroup mapping of a fiducial state $\rho$ into
a state $\Gamma(\rho)$ with a number of thermal photons
"added". 
Generalizing an idea from \cite{hall} we shall define a Gaussian noise map 
$\Gamma : \rho \rightarrow \Gamma(\rho) $ for any density 
operator $\rho$ in the following way:
\begin{equation}
\Gamma(\rho) = \int p_{{\cal G}}(v) V(v) \rho V(-v) dv 
\end{equation}
where $p_{{\cal G}}(v)$ is a probability distribution 
on the phase space of Gaussian form:
\begin{equation}
p_{{\cal G}}(v)= \pi^{-n} \sqrt{det{\cal G}} \exp\{-v^T{\cal G}v \}
\end{equation}
(${\cal G}$ is a positive definite $2n \times 2n$ matrix)
and $V(u)$ are the Weyl operators.
It is easy to see that in the case when $\rho$ is a quasifree
state with the characteristic function
\begin{equation}
CF_{u}(\rho) = \exp\{-{u^TAu \over 4}\} 
\end{equation}
the characteristic function of the state $\Gamma(\rho)$
is given by
\begin{equation}
CF_{u}(\Gamma(\rho)) = \exp\{-{u^T(A-J{\cal G}^{-1}J)u \over 4}\}
\end{equation}
In the following we shall use the notation
$\Gamma(A) = A - J{\cal G}^{-1}J$.
In the general case ${\cal G}$ must be of the following
form ${\cal G}= O_{{\cal G}}^T\left(\matrix{D_{{\cal G}}
&0 \cr 0&D_{{\cal G}}\cr} \right)O_{{\cal G}}$ where 
$O_{{\cal G}}$ is an orthogonal symplectic matrix.
Hence 
\begin{equation}
\Gamma(A)= A + O_{{\cal G}}^T\left(\matrix{D_{{\cal G}}^{-1}
&0 \cr 0&D_{{\cal G}}^{-1}\cr} \right)O_{{\cal G}}.
\end{equation}

\subsection{The Gaussian noise in the one mode case}
Let us concentrate on the one mode case
when $A= dS^TS$ where $S=O{\cal M}O^{'}$ with
$O$ and $O^{'}$ orthogonal matrices and ${\cal  M}=
\left (\matrix{m & 0\cr 0 &{1 \over m}\cr} \right)$.
Then it is evident that $Tr A=d(m^2+{1 \over m^2})$
and $det A=d^2$. Hence the squeezing parameter $m$
is determined by the invariants $det A$ and $TrA$ 
of the correlation matrix $A$ from the equation
\begin{equation}
(m^2+{1 \over m^2})= {Tr A \over \sqrt{detA}}
\end{equation}
Evidently, to the correlation matrix $\Gamma(A)$
there corresponds a new squeezing parameter 
$\Gamma(m)$ given by the analogous equation with
$A$ replaced with $\Gamma(A)$.
From the formula (26) it follows that 
\begin{equation}
det\Gamma(A)= det(g^{-1}I+ d(O_{{\cal G}}^T)^{-1}S^TSO_{{\cal G}}^{-1}) 
\end{equation}
and
\begin{equation}
Tr\Gamma(A) = Tr {\cal G}^{-1}+ TrA
\end{equation}
From the formula (28) we obtain that
\begin{equation}
det\Gamma(A) = det{\cal G}^{-1}+{Tr{\cal G}^{-1}TrA 
\over 2}+detA
\end{equation}
Therefore the equation for the squeezing
parameter $\Gamma(m)$ is
\begin{equation}
\Gamma(m)^2+{1 \over \Gamma(m)^2}= {Tr {\cal G}^{-1}+ TrA
\over {det{\cal G}^{-1}+{Tr{\cal G}^{-1}TrA 
\over 2}+detA}}
\end{equation}
With the above parametrization we have
\begin{equation}
\Gamma(m)^2= \sqrt{{{1 \over g}+dm^2 \over {{1 \over g}+{d \over m^2} }}}
\end{equation}
Also we have
\begin{equation}
\Gamma(d)=  \sqrt{({1 \over g}+dm^2)({1 \over g}+{d \over m^2})}
\end{equation}
and
\begin{equation}
g^{-1}= {d[(m^2+{1 \over m^2})+\sqrt{(m^2-{1 \over m^2})^2+
4({\Gamma(d) \over d})^2} \over 2} 
\end{equation}
A direct relation between $\Gamma(m)$ and $\Gamma(d)$ is the
following:
\begin{equation}
\Gamma(m)=\sqrt{{m^2-{1 \over m^2} \over 2{\Gamma(d) 
\over d}}+\sqrt{1+({m^2-{1 \over m^2} \over 2{\Gamma(d) 
\over d}})^2}}
\end{equation}

\subsection{Comparison with the previous results}
In order to compare the above results with those from the papers
\cite{mussli,pmtm} the following identifications are made:
$g={1 \over 2 \bar{n}}$, $m^{'}=\Gamma(m)$, \c si
$\Gamma(d=1)=2\bar{n}^{'}+1$ and $m=\exp(r)$.
Then the first two equations (A6) from the Appendix A
of the paper \cite{mussli} become:
\begin{equation}
m^2-{1 \over m^2}= -\Gamma(1)(\Gamma(m)^2-{1 \over \Gamma(m)^2})
\end{equation}
and
\begin{equation}
{1 \over 2g}+ {1 \over 4}(m^2+{1 \over m^2})= {\Gamma(1) \over 4}
(\Gamma(m)^2+{1 \over \Gamma(m)^2})
\end{equation}
The minus sign from the equation (36) is not correct .
Wihout this sign the equation (36) becomes:
\begin{equation}
\sqrt{{1+gm^2 \over 1+{g \over m^2}}}-\sqrt{{1+{g \over m^2} \over 1+gm^2}}=
{g(m^2-{1 \over m^2}) \over \sqrt{(1+gm^2)(1+{g \over m^2})}}
\end{equation}
and is evidently fullfiled.
The equation (37) becomes:
\begin{equation}
{({1 \over g}+m^2)+({1 \over g}+{1 \over m^2}) \over
\sqrt{({1 \over g}+m^2)({1 \over g}+{1 \over m^2})}}=
\sqrt{{ {1 \over g}+m^2 \over {1 \over g}+{1 \over m^2}}}
+\sqrt{{{1 \over g}+{1 \over m^2} \over {1 \over g}+m^2 }}
\end{equation}
and is also evidently fullfiled.

In the general case $1 < \Gamma(m) \leq m$, i.e.
the Gaussian noise map reduces the squeezing
but it does not suppress it.
Also $ \Gamma(d) \geq d$ i.e. the
Gaussian noise map increases the number of thermal
photons.
A nonclassical state $\rho$
becomes a classical one under the map $\Gamma$
when $\Gamma(d) > \Gamma(m)^2$,
i.e. when ${1 \over g} + {d \over m^2} > 1$.  
This inequality is valid for any values of $d$ and
$m$ when $g \leq 1$, which is evidently fullfiled.

\section{The Holevo distance}
The characteristic function of a multimode squeezed thermal
state is given by:
\begin{equation}
CF_{u}(\rho)= \exp\{-{1 \over 4}u^TAu\}
\end{equation}
where $A=2\Sigma$ and $\Sigma$ is the correlation matrix
of the state.
We have for any two density operators $\rho_{1}$
and $\rho_{2}$:
\begin{equation}
CF_{u}(\rho_{1}\rho_{2})= (2\pi)^{-n} \int CF_{v+{u \over2}}(\rho_{1})
CF_{{u \over 2}-v}(\rho_{2}) \exp\{{iv^TJu \over 2}\}dv
\end{equation}
where $J=\left(\matrix{~0&I \cr -I&0 \cr} \right)$.
This becomes in the particular case when $\rho_{1}=\rho_{2}=
\sqrt{\rho}$:
\begin{equation}
CF_{u}(\rho)=(2\pi)^{-n}\int CF_{{u \over 2}+v}(\sqrt{\rho})
CF_{{u \over 2}-v}(\sqrt{\rho})\exp\{{iv^TJu \over 2}\}dv
\end{equation}
If we suppose that
\begin{equation}
CF_{u}(\sqrt{\rho})=K \exp\{-{1 \over 4}u^T\phi(A)u\}
\end{equation}
then it follows that
\begin{equation}
CF_{u}(\rho)=K^2 (det\phi(A))^{-{1 \over 2}} \exp\{-{1 \over 8}
u^T(\phi(A)-J\phi(A)^{-1}J)u\}
\end{equation}
In order that this equality be valid for all values of $u$
we must have $K=(det\phi(A))^{{1 \over 4}}$ and
\begin{equation}
\phi(A)-J\phi(A)^{-1}J=2A
\end{equation}
We can put this equation in the following form
\begin{equation}
J\phi(A)J\phi(A)-I=2J\phi(A)JA
\end{equation}
Now we shall prove that this equation has a solution
which is given by
\begin{equation}
\phi(A) = A(I+\sqrt{I+(JA)^{-2}})
\end{equation}
\bigskip
Indeed, we have
\begin{equation}
A = S^T{\cal D}S
\end{equation}
with $S^TJS=J$ (i.e. $S$ is a symplectic matrix)
and ${\cal D} \geq I$ is a diagonal matrix.
It is well known that if $S$ is a symplectic 
matrix then $S^T$ and $S^{-1}$ are also symplectic
matrices. From this fact we obtain that
\begin{equation}
(JA)^{-2} = -S^{-1}{\cal D}^{-2}S
\end{equation}
Also we have
\begin{equation}
J\phi(A)JA= -S^{-1}{\cal D}({\cal D}+\sqrt{{\cal D}^2-I})S
\end{equation}
and
\begin{equation}
J\phi(A)J\phi(A)= -S^{-1}({\cal D}+\sqrt{{\cal D}^2-I})^2S
\end{equation}
Because $({\cal D}+\sqrt{{\cal D}^2-I})^2+I=2{\cal D}
({\cal D}+\sqrt{{\cal D}^2-I})$
the desired result follows.
The following form of the function $\phi(A)$ is also useful:  
\begin{equation}
\phi(A)= -S^{-1}({\cal D}+\sqrt{{\cal D}^2-I})S
\end{equation}
It is interesting to point out that in the case of a pure state
${\cal D}=I$ and $\phi(A)=A$.
Now we can calculate $Tr\sqrt{\rho_{1}}\sqrt{\rho_{2}}$.
From the general formula:
\begin{equation}
TrB_{1}B_{2}= (2\pi)^{-n} \int CF_{u}(B_{1})
CF_{-u}(B_{2})du
\end{equation}
it follows that
\begin{equation}
Tr\sqrt{\rho_{1}}\sqrt{\rho_{2}}=(2\pi)^{-n} \int 
CF_{u}(\sqrt{\rho_{1}})
CF_{-u}(\sqrt{\rho_{2}})du
\end{equation}
It is easy to compute this integral. The result is
\begin{equation}
Tr\sqrt{\rho_{1}}\sqrt{\rho_{2}} = \sqrt{{\sqrt{det \phi(A_{1})
det \phi(A_{2})} \over det({\phi(A_{1})+\phi(A_{2}) \over 2})}}
\end{equation}
When $\rho_{2}$ is a squeezed state (i.e. when
$A_{2}= O_{2}^T \left(\matrix{M^2&0 \cr 0&M^{-2}}
\right)O_{2}$)
it is plaussible to suppose that the maximum value
of this quantity is obtained for $D_{1}=I$,
and $O_{1}=O_{2}$. In this case 
\begin{equation}
\chi(\rho)={ 1 \over det({M+M^{-1} \over 2})}
\end{equation}
(where we have denoted the density matrix $\rho_{2}$
by $\rho$).
It is clear from this formula that the nonclassicity of
the state $\rho$ is entirely due to the squeezing.
When $M=I$ we have no squeezing and $\chi(\rho)=1$.

\subsection{The one-mode case}
In the one-mode case
the most general classical state is given by a characteristic
function with $A_{1}=O_{1}^T \left(\matrix{d_{1}m_{1}^2&0 \cr 0&{d_{1}
\over m_{1}^2}\cr} \right)O_{1}$, 
where $O^TO=I$ and the positivity of $A_{1}-I$ requires 
the validity of the inequalities $d_{1} \geq m_{1}^2$ and
$d_{1} \geq m_{1}^{-2}$ which are not independent. Wnen
one of them is satisfied then the other is also satisfied.
Because for any symplectic matrix $S$ we have
$detS=1$ it follows that 
$det\phi(A_{k})=(d_{k}+\sqrt{d_{k}^2-1})^2$ for $k=1,2$.
If $\rho_{2}$ is also a classical state then
\begin{equation}
\sup_{\rho_{1}}Tr\sqrt{\rho_{1}}\sqrt{\rho_{2}}=1
\end{equation}
is obtained for $\rho_{1}=\rho_{2}$.

When $\rho_{2}$ is a squeezed state (i.e. when
$A_{2}= O_{2}^T \left(\matrix{d_{2}m_{2}^2&0 
\cr 0&{d_{2} \over m_{2}^2}\cr} \right)O_{2}$)
then we have:
\begin{equation}
Tr\sqrt{\rho_{class}}\sqrt{\rho}={2 \over \sqrt{{(\phi(d_{1})-\phi(d_{2}))^2
\over \phi(d_{1})\phi(d_{2})} + F(\Delta\theta ,m_{1}, m_{2})}}
\end{equation}
where we have denoted with $F$ the following function
\begin{eqnarray}
&&
\nonumber
F(\Delta\theta ,m_{1}, m_{2})=\\
&&
\nonumber
2+sin(\Delta\theta)^2(m_{1}^2m_{2}^2+
{1 \over m_{1}^2m_{2}^2})+cos(\Delta\theta)^2(({m_{1} \over m_{2}})^2+
({m_{2} \over m_{1}})^2)\\
\end{eqnarray}
The maximum value of this function corresponds to the minimum
value of the fucntion under the square-root:
\begin{equation}
H(d_{1}, d_{2}, \Delta\theta, m_{1}, m_{2})=
{(\phi(d_{1})-\phi(d_{2}))^2
\over \phi(d_{1})\phi(d_{2})} + F(\Delta\theta ,m_{1}, m_{2})
\end{equation}
It is evident that the minimum value of $H$
is attained for $d_{1}=d_{2}$ and for those values
of $\Delta\theta$, $m_{1}$ and $m_{2}$ which minmize
the function $F$. The minimum value of the function $F$ is
equal with $2$ and is attained either for $\Delta\theta =0$,
and $m_{1}=m_{2}$ or for $\Delta\theta={\pi \over 2}$, 
and $m_{1}=m_{2}^{-1}$. But when $\rho_{2}$ is not a classical
state, then $d_{1}=d_{2} < m_{2}^2=m_{1}^2$  or
$d_{1}=d_{2} > m_{2}^{-2}=m_{1}^{-2}$ in the first case
and $d_{1}=d_{2} < m_{2}^2=m_{1}^{-2}$ or
$d_{1}=d_{2} > m_{2}^{-2}=m_{1}^2$ in the second case. 
In all these situations
the conditions for the classicality of the state $\rho_{1}$
are not satisfied.
Because the function $F$ is monotonely decreasing for 
$m_{1} \leq \sqrt{d_{2}} < m_{2}$ or for $m_{1} \geq {1 \over
\sqrt{d_{2}} }> m_{2}$ it follows that the minimum
value of $F$ is in both cases equal with $({\sqrt{d_{2}}
\over m_{2}}+ {m_{2} \over \sqrt{d_{2}}})^2$.
Hence we have obtained that the nonclassical distance
for a thermal squeezed state $\rho$:
\begin{equation}
\chi(\rho) = {2 \over {\sqrt{d} \over m} + {m \over
\sqrt{d}}}
\end{equation}

\subsection{The increase of nonclassical distance under
Gaussian noise}
The nonclassical
distance $\chi(\Gamma(\rho))$ in the case of
a one mode thermal squeezed state $\rho$
is then given by
\begin{equation}
\chi(\Gamma(\rho))= {2 \over {\sqrt{\Gamma(d)} \over \Gamma(m)}+
{\Gamma(m) \over \sqrt{\Gamma(d)}}}
\end{equation}
when $\Gamma(d) > \Gamma(m)^2$
or by $\chi(\rho)=1$ when $\Gamma(d) < \Gamma(m)^2$.
The intuitive fact according to which $\Gamma(\rho)$
is closer to a classical state than $\rho$ is reflected
quantitatively in the inequality
\begin{equation}
\chi(\Gamma(\rho)) \geq \chi(\rho).
\end{equation}
Hence we must prove that
\begin{equation}
{\Gamma(m)\sqrt{\Gamma(d)}  \over m\sqrt{d}} \geq
{\Gamma(d) + \Gamma(m)^2 \over d + m^2}
\end{equation}
or in a more convenient form
\begin{equation}
{\Gamma(m)^2{\Gamma(d)}  \over (\Gamma(d) + \Gamma(m)^2)^2}
\geq {{d \over m^2} \over ({d \over m^2}+ 1)^2}
\end{equation}
which becomes
\begin{equation}
{({d \over m^2}+ {1 \over g}) \over ({d \over m^2}+ {1 \over g}+1)^2}
\geq  {{d \over m^2} \over ({d \over m^2}+ 1)^2}
\end{equation}
From this it follows that the inequality
$\chi(\Gamma(\rho)) \geq \chi(\rho)$ is valid
only for ${1 \over g} \leq  {m^2 \over d}-{d \over m^2}$.
This limitation on the number of thermal photons $2{\bar n}=
{1 \over g}$ introduced
by the Gaussian noise map is inacceptable.

\section{The fidelity distance}

The fidelity $F(\rho_{1},\rho_{2})$ for two density operators
$\rho_{1}$ and $\rho_{2}$ is defined by
\begin{equation}
F(\rho_{1},\rho_{2})=
Tr\left(\sqrt{\sqrt{\rho_{1}}\rho_{2}\sqrt{\rho_{1}}}\right)^2.
\end{equation}
Since the characteristic function of a 
product of operators whose characteristic functions are Gaussians
is also a Gaussian and 
the characteristic function of the square root of a Gaussian density operator
is a Gaussian we can find a simple 
formula for the characteristic function of the operator $\sqrt{\rho_{1}}
\rho_{2}\sqrt{\rho_{1}}$:  
\begin{equation}
CF_{z}(\sqrt{\rho_{1}}\rho_{2}\sqrt{\rho_{1}})=\sqrt{L}\exp\left\{-{1 \over 4}
z^T{\cal O}z\right\},
\end{equation}
where
\begin{eqnarray}
&&
\nonumber
L^{-1}=det\Phi(A_{1})^{-1} 
det\left({\Phi(A_{1})+A_{2} \over 2}\right)\\
&&
\nonumber
det\left({A_{2}+\Phi(A_{1})-{\cal U}
\over 2}\right)\\
\end{eqnarray}
where ${\cal U}=(A_{2}-iJ)(\Phi(A_{1})+A_{2})^{-1}(A_{2}+iJ)$,
and
\begin{eqnarray}
&&  
\nonumber
{\cal O}= \Phi(A_{1})-(\Phi(A_{1})-iJ)[A_{2}+\Phi(A_{1})-\\
&&
\nonumber
(A_{2}-iJ)(\Phi(A_{1})+A_{2})^{-1}(A_{2}+iJ)]^{-1}(\Phi(A_{1})+iJ).
\end{eqnarray}
Then applying the result of the preceeding section
we can obtain the characteristic function of
$\sqrt{\sqrt{\rho_{1}}\rho_{2}\sqrt{\rho_{1}}}$,
\begin{eqnarray}
&&
\nonumber
CF_{z}\left(\sqrt{\sqrt{\rho_{1}}\rho_{2}\sqrt{\rho_{1}}}\right)=\\
&&
\nonumber
\left[Ldet\Phi({\cal O})\right]^{{1 \over 4}}\exp\left\{-{1 \over 4}z^T\Phi
({\cal O})z\right\}.\\
\end{eqnarray}
From this formula and the property 1 of the characteristic function
we obtain
\begin{equation}
F(\rho_{1},\rho_{2})=\sqrt{Ldet\Phi({\cal O})}. 
\end{equation}
We remark that 
\begin{equation}
det\Phi({\cal O})=det{\cal O}det\left[I+\sqrt{I+(J{\cal O})^{-2}}\right].
\end{equation}
In order to simplify the formula for the fidelity we observe that
\begin{eqnarray}
&&
\nonumber
t_{ijk}=Tr\rho_{i}\rho_{j}\rho_{k}=
det\left({A_{i}+A_{j} \over 2}\right)\\
&&
\nonumber
det\left[{A_{j}+A_{k}-(A_{j}-iJ)
(A_{i}+A_{j})^{-1}(A_{j}+iJ) \over 2}\right],
\end{eqnarray}
and that $t_{123}=t_{231}=t_{312}$.
If we take in this last identity $\Phi(A_{1})$
instead of $A_{1}$ we obtain
\begin{eqnarray}
&&
\nonumber
det\left[{\Phi(A_{1})+A_{2} \over 2}\right]\\
&&
\nonumber
det\left[{A_{2}+\Phi(A_{1})-{\cal U} \over 2}\right]\\
&&
\nonumber
=det\left({A_{1}+A_{2} \over 2}\right)det\Phi(A_{1}).
\end{eqnarray}
Hence we get
\begin{equation}
L=\left[det\left({A_{1}+A_{2} \over 2}\right)\right]^{-1}.
\end{equation}

\subsection{The one mode case}

In \cite{scut1} we have obtained an expression  for the fidelity
in the one mode case. This formula can be reobtained as a consequence of
the above general formula. In the one mode case all matrices
are $2 \times 2$ matrices. For a $2 \times 2$ matrix ${\cal O}$
we have
\begin{equation}
\Phi({\cal O})= \epsilon {\cal O},
\end{equation}
where $\epsilon=1+\sqrt{1-{1 \over det{\cal O}}}$ and
$det\Phi({\cal O})=(\sqrt{det{\cal O}}+\sqrt{\det{\cal O}-1})^2$.
From these considerations it follows that
\begin{equation}
F(\rho_{1},\rho_{2})={2 \over \sqrt{det(A_{1}+A_{2})}
(\sqrt{det{\cal O}}-\sqrt{det{\cal O}-1})}.
\end{equation}
Thus  it is sufficient to compute $det{\cal O}$. We shall
denote by ${\cal P}$ the product $(detA_{1}-1)(detA_{2}-1)$.
After simple but long computations we obtain
\begin{equation}
det{\cal O}= 1 + {{\cal P} \over det(A_{1}+A_{2})},
\end{equation}
which gives the result of \cite{scut1} 
\begin{equation}
F(\rho_{1},\rho_{2})=
{2 \over \sqrt{det(A_{1}+A_{2})+
{\cal P} }-\sqrt{{\cal P}}}.
\end{equation}
With the parametrization taken in the subsection 4.1.
we have:
\begin{eqnarray}
&&
\nonumber
F(\rho_{1},\rho_{2})=\\
&&
\nonumber
{2 \over \sqrt{d_{1}^2d_{2}^2+1+d_{1}d_{2}[F(\Delta \theta,m_{1},m_{2})-2] 
}-\sqrt{(d_{1}^2-1)(d_{2}^2-1)}}.\\
\end{eqnarray}
Then
\begin{equation}
\phi(\rho)= \sup_{\rho_{cl}}F(\rho_{cl},\rho)=
{2 \over \sqrt{(d^2-1)^2+d^2[{\sqrt{d} \over m}+{m \over \sqrt{d}}]^2} 
-(d^2-1)}.
\end{equation}
We stress the fact that for nonclassical Gaussian states we have
$\sqrt{d} < m$.
The fact that $\phi(\Gamma(\rho)) \geq \phi(\rho)$ is a direct consequence
of the definition of the Gaussian noise map and of the property
7 of the fidelity (transition probability) given in the introduction.

\section{Conclusions}
We have considered the problem of nonclassical distance from the point
of view of distinguishability between quantum states. In the
case of Gaussian states, using an explicit formula for the fidelity,
we have obtained an explicit formula for the nonclassical distance.
The Gaussian noise was used to eliminate an attractive candidate
for the definition of nonclassical distance. In the particular case
of pure Gaussian states our results are comparable with the
upper bounds obtained for Hillery's nonclassical distance \cite{hill1},
which is defined using the trace norm, because these upper bounds
contain the overlaps between the squeezed states and the
coherent states.

\end{document}